\documentclass[a4paper,11pt]{article}
\pdfoutput=1 

\usepackage{jheppub} 
\usepackage{empheq}
\usepackage{bm}
\usepackage{slashed}
\usepackage[T1]{fontenc} 
\usepackage[all]{xy}
\usepackage{rotating}
\usepackage{amsmath} 
\usepackage{mathtools}

\usepackage{slashed}

\def\a{\alpha}

\def\CA{{\cal A}}

\def\CO{{\cal O}}

\def\beq#1\eeq{\begin{align}#1\end{align}}

\preprint{SNUTP20-002}
\title{\boldmath Black holes and Large $N$ complex saddles in 3D-3D correspondence }

\abstract{We study the large $ N$ sign oscillation  of the twisted indices for 3D   theories of class $\mathcal{R}$ obtained from  M5-branes wrapped on a hyperbolic 3-manifold.  Holographically, the oscillatory behavior can be understood from the imaginary part of  on-shell  actions for the two Euclidean supergravity solutions, Bolt$_{\pm}$ with $p=0$, which are Wick rotation of magnetically charged   AdS$_4$ black holes.   The two solutions have the same imaginary part with opposite sign. The imaginary part comes from the $F\wedge F$-term in the supergravity and the coefficient is   proportional to the Chern-Simons invariant of  3-manifold.  Combining the holographic computation with 3D-3D relation for twisted indices, we propose a non-trivial mathematical conjecture regarding the phase factor of a twisted Reidemeister-Ray-Singer torsion on  hyperbolic 3-manifold.}

\author[a]{Sunjin Choi,}
\author[b,c]{Dongmin Gang,}
\author[d,e]{Nakwoo Kim}

\affiliation[a]{Department of Physics and Astronomy \& Center for Theoretical Physics,\\Seoul National University, Seoul 08826, Republic of Korea}

\affiliation[b]{Asia Pacific Center for Theoretical Physics (APCTP),
	Pohang 37673, Korea}

\affiliation[c]{Department of Physics, Pohang University of Science and Technology (POSTECH), \\Pohang 37673, Republic of Korea}

\affiliation[d]{Department of Physics and Research Institute of Basic Science, Kyung Hee University, \\26 Kyungheedae-ro, Dongdaemun-gu, Seoul 02447, Republic of Korea
}

\affiliation[e]{School of Physics, Korea Institute of Advanced Study, \\ 85 Hoegi-ro, Dongdaemun-gu, Seoul 02455, Republic of Korea
}
\emailAdd{csj37100@snu.ac.kr}

\emailAdd{dongmin.gang@apctp.org}
\emailAdd{nkim@khu.ac.kr}

\begin{document} 
\maketitle
\flushbottom

\section{Introduction and summary}
AdS/CFT correspondence provides a consistent non-perturbative definition of quantum gravity in asymptotically AdS spacetimes  \cite{Maldacena:1997re}. The correspondence is  expected to provide a systematic way to understand microscopic physics of black holes, which are key research objects in quantum gravity. One immediate and important application is to statistically account for the Bekenstein-Hawking formula of the black hole entropy from the field theory. After an early attempt using superconformal index in \cite{Kinney:2005ej},  it has long been believed that the index cannot account for the exponentially growing behavior of the microstates of black hole due to huge cancellation between bosonic/fermionic states. 
The situation began to change after the works of \cite{Benini:2015eyy,Benini:2016hjo}, where  they obtained an exponentially growing behavior from another type of BPS index called topologically twisted indices \cite{Benini:2015noa,Benini:2016hjo,Closset:2016arn}. The exponential growth at leading order in  $1/N$ expansion nicely matches with the expected behavior from Bekenstein-Hawing entropy formula of  dyonic black holes. The holographic computation was done using the AdS$_4$/CFT$_3$ associated to the M2-branes and after that the work was extended to various types of AdS$_4$/CFT$_3$ examples. Finally, the entropies of electrically charged rotating AdS black holes were also reproduced \cite{Choi:2018hmj,Benini:2018ywd,Choi:2019zpz,Choi:2019miv} from  a large $N$ limit on superconformal index of dual   field theory by revisiting the computation in \cite{Kinney:2005ej}.    More recently, it was shown that one can study not only the absolute degeneracy saturating the Bekenstein-Hawking entropy from the superconformal index  but also the sign oscillation of the index through the complex conjugate pair of large $N$ saddle points of the Legendre transformation of the free energy \cite{Agarwal:2020zwm}. In the dual gravity side, these conjugate large $N$ saddles correspond to the complex conjugate pair of the Euclidean solutions, which are the Wick rotated version of the electrically charged rotating  AdS$_5$ BPS black hole solutions \cite{Cabo-Bizet:2018ehj,Cassani:2019mms}.

In this paper, we study the sign oscillation of (2+1)D topologically twisted indices, which holographically count, with sign, microstates of  magnetically charged AdS$_4$ black holes.  For the purpose, we study a particular class of  AdS$_4$/CFT$_3$ correspondence arising from $N$ M5-branes wrapped on a compact hyperbolic 3-manifold $M$. The superconformal field theories are often called 3D theories of class $\mathcal{R}$.
 One nice feature of these 3D superconformal field theories is that we can use the 3D-3D correspondence to compute various supersymmetric quantities from $SL(N, \mathbb{C})$ Chern-Simons theory invariants  of $M$ \cite{Dimofte:2010tz,Terashima:2011qi,Dimofte:2011ju}. In \cite{Gang:2019uay}, using the 3D-3D correspondence, the large $N$ limit of the topologically twisted index $I_g$ on Riemann surface $\Sigma_g$  of genus $g>1$  was computed and its large $N$ exponential growth was shown to agree with the Bekenstein-Hawking entropy of magnetic BPS black holes in the AdS$_4$ dual with AdS$_2\times \Sigma_g$ near-horizon geometry \cite{Romans:1991nq,Caldarelli:1998hg}. Standard lore is that bosonic states always dominate in the microstates of the magnetic BPS black holes and thus the twisted indices are always positive at sufficiently large $N$ \cite{Azzurli:2017kxo}. 
In contrast, we will see that there exists sign oscillation in the twisted indices of class $\mathcal{R}$ theories at large $N$. The  sign oscillation can be studied in two independent ways.  First, using the 3D-3D correspondence, we demonstrate that the large $N$ limit of the twisted index receives contributions from a pair of the equally dominant complex conjugate saddles, exhibiting the  sign oscillation as $N$ or $g$ varies. Such sign oscillation is governed by the Chern-Simons invariant of the hyperbolic 3-manifold $M$, $cs(M)$, as following:
\begin{equation} \label{osc-ind}
\begin{aligned}
I_g [T_N[M]] &\sim \exp \left( (g-1)\frac{\textrm{vol}(M)+i \, cs (M)}{3\pi} N^3 + \mathcal{O}(N) \right) + (c.c.) \\
& = \exp \left( (g-1)\frac{\textrm{vol}(M)}{3\pi} N^3 + \mathcal{O}(N) \right) \cos \left( (g-1)\frac{cs (M)}{3\pi} N^3 + \mathcal{O}(N) \right)\ .
\end{aligned}
\end{equation}
Here, $\textrm{vol}(M)$ is the hyperbolic volume. The exponentially growing factor equals the Bekenstein-Hawking entropy of the magnetically charged BPS black holes in the AdS$_4$ dual, i.e. $S_{\textrm{BH}} = (g-1)\frac{\textrm{vol}(M)}{3\pi} N^3$ \cite{Gang:2018hjd}.
Next, we compute the (regularized) Euclidean on-shell action in the AdS$_4$ gravity dual. After the Wick rotation of the magnetic BPS black hole, the Euclidean solutions come in a pair; the Bolt$_\pm$ solutions with $p=0$ and $g>1$ \cite{Toldo:2017qsh}. While Bolt$_\pm$ solutions are purely real, we show that their contributions to the Euclidean on-shell action are complex conjugate to each other. The imaginary part of the on-shell action comes from the topological $\theta$-term in 4D $\mathcal{N}=2$ minimal gauged supergravity whose bosonic action in the Euclidean signature reads
\begin{equation} \label{SUGRA-action}
\begin{aligned}
S = - \frac{1}{16\pi G_4} \int d^4x \sqrt{g}\left( R + \frac{6}{L^2} - F_{\mu\nu}F^{\mu\nu} \right)  + i \frac{\theta}{4\pi^2L^2} \int F \wedge F \, ,
\end{aligned}
\end{equation}
where $G_4$ is the 4D Newton's constant and $L$ is the AdS$_4$ radius. Such purely imaginary $\theta$-term is often overlooked in the literature. The above 4D action can be derived from the consistent KK truncation of 7D maximal gauged supergravity on $M$ \cite{Pernici:1984xx,Gauntlett:2000ng,Donos:2010ax}. Also, the 7D gauged supergravity can be uplifted to 11D supergravity on a squashed 4-sphere $\tilde{S}^4$ \cite{Nastase:1999cb,Nastase:1999kf,Cvetic:2000ah}. As the 3D class $\mathcal{R}$ theory is holographically dual to the M-theory on AdS$_4 \times M \times \tilde{S}^4$, such successive embedding relates the macroscopic parameters $L, \, G_4, \, \theta$ to the microscopic parameters $N, \, \textrm{vol}(M), \, cs(M)$ as following:
\begin{equation} \label{para-map}
\frac{L^2}{G_4} = \frac{N^3+\mathcal{O}(N)}{6} \frac{4 \,\textrm{vol}(M)}{\pi^2}\ , \quad \theta = -\frac{N^3+\mathcal{O}(N)}{6} \frac{2 \, cs(M)}{\pi}\ .
\end{equation}
Here, $\CO(N)$ refers to a quantum shift, which is difficult to determine from the supergravity analysis. With these relations and appropriate counterterms needed for the holographic renormalization \cite{Azzurli:2017kxo,Toldo:2017qsh}, the Euclidean on-shell action evaluated on Bolt$_\pm$ with $p=0$ are given by
\begin{equation}
S[\textrm{Bolt}_{\pm}] = (1-g) \frac{N^3+ \mathcal{O}(N)}{6}\frac{2}{\pi}(\textrm{vol}(M)\pm i \, cs (M))\ ,
\end{equation}
which is consistent with the field theory computation \eqref{osc-ind} at the leading $N^3$ order. Here, $\textrm{vol}(M)\pm i \, cs (M)$ is the complex volume, which is the holomorphic topological invariant of the hyperbolic 3-manifold $M$ \cite{neumann1999bloch}. The invariant is well-defined only modulo $i \pi^2$. For the exponentiated classical action $e^{-S[{\rm Bolt}_{\pm}]}$  to be well-defined,  the undetermined quantum  shift $\CO(N)$ should satisfy following condition:
\begin{align}
\frac{N^3+\mathcal{O}(N)}{6}  \in \mathbb{Z}\;.
\end{align}
Note that the  $\theta$-term in \eqref{SUGRA-action} does not appear from the consistent truncation of 11D supergravity on the Sasaki-Einstein 7-manifold (SE$_7$). That is the reason why we did not have to consider such term when we analyze the black holes in AdS$_4 \times \textrm{SE}_7$ dual to M2-brane SCFTs \cite{Toldo:2017qsh}. Also, the on-shell value of the $\theta$-term vanishes if there is no topological twist on AdS$_4$, so it does not affect the on-shell action of the electrically charged rotating BPS black holes in AdS$_4$ regardless of embedding in 11D supergravity \cite{Cassani:2019mms}.

The rest of this paper is organized as follows. In section \ref{sec : review on complex saddles}, we review complex saddles and  sign oscillation of  4D superconformal index in the large $N$ limit.  In section \ref{sec : complex saddles in 3D-3D}, we study the sign oscillation in the large $N$ twisted indices  of 3D class $\mathcal{R}$ theories from  both of  3D-3D dual complex Chern-Simons theory and
 holographic dual supergrvity. As a byproduct of the analysis, we propose a concrete mathematical  conjecture in \eqref{conjecture},  which is numerically confirmed in Appendix \ref{App : numerical check}.

\section{Large $N$ complex saddles in AdS/CFT}  
\label{sec : review on complex saddles}
In this section, we briefly review basic ideas about the large $N$ complex saddle points \cite{Choi:2018hmj,Choi:2018vbz,Choi:2019miv,Agarwal:2020zwm,Cabo-Bizet:2018ehj,Cassani:2019mms} associated to  the electrically charged rotating AdS black holes.   
In the dual conformal field theory side, the microstates of black holes correspond to the local operators with fixed macroscopic charges.

\subsection{Conformal Field Theory side}
Following \cite{Agarwal:2020zwm}, our concrete example is the superconformal index \cite{Romelsberger:2005eg,Kinney:2005ej} of 4D $\mathcal{N}=4$ supersymmetric-Yang-Mills theory with $U(N)$ gauge group, which is dual to the type IIB string theory on AdS$_5 \times S^5$. The (unrefined) superconformal index is defined as
\begin{equation}
Z= \textrm{Tr}_{\mathcal{H}_{\rm rad}(S^3)} [(-1)^F x^{3(R+2J)}] = \sum_{j=0}^\infty \Omega_j x^j\ ,
\end{equation}
where $R$ is the superconformal R-charge, $J=\frac{J_1+J_2}{2}$ with $J_i$'s being two angular momenta on $S^3$, and $j\equiv 3(R+2J)$. The trace is taken over the radially quantized Hilbert space $\mathcal{H}_{\rm rad}(S^3)$ on $S^3 \times \mathbb{R}$ whose states correspond to local operators on $\mathbb{R}^4$. The above index receives contributions only from the BPS states with $ j \geq 0$ and $\Omega_j$ counts the $(-1)^F$-weighted number of BPS states with the charge $j$. In the large $N$ limit, one may compute the free energy $\log Z \sim \mathcal{O}(N^2)$ using the saddle point approximation, where the large $N$ saddle points are supposed to correspond to the black hole saddle points in the AdS$_5$ gravity dual. In order to extract the entropy from the large $N$ free energy, one should perform the Legendre transformation to the microcanonical ensemble, which is the saddle point approximation of the inverse Laplace transformation
\begin{equation}
\Omega_j = \frac{1}{2\pi i}\oint \frac{dx}{x} x^{-j} Z(x)
\end{equation}
at macroscopic charge $j \sim \mathcal{O} (N^2)$. A known fact is that the dominant saddle point values $x_*$ of $x$ are complex at real positive charge $j$. Naively, this yields the complex entropy $S(j)$ at real charge. However, as pointed out in \cite{Choi:2019zpz,Agarwal:2020zwm}, the unitarity of our QFT guarantees that there always exists a complex conjugate saddle point for any such complex saddle point. Then, adding two equivalently dominant contributions from the complex conjugate pair of saddle points, we obtain
\begin{equation}
\Omega_j \sim e^{S(j)} + e^{\overline{S(j)}} \sim \exp [\textrm{Re} (S(j)) + \cdots] \cos [\textrm{Im} (S(j)) + \cdots]\ , \label{Oscillatory for electrical BHs}
\end{equation}
where $\cdots$ denote possible subleading corrections in the large $N$ expansion. In \cite{Agarwal:2020zwm,Murthy:2020rbd}, they numerically studied the index of $\mathcal{N}=4$ SYM, and in  \cite{Agarwal:2020zwm} it was shown that the above formula is matched very nicely by the numerically computed index at reasonably large $N$ and $j$. The first macroscopic exponential factor captures the Bekenstein-Hawking entropy of electrically charged rotating BPS black holes in AdS$_5 \times S^5$ \cite{Gutowski:2004ez,Gutowski:2004yv,Chong:2005da,Kunduri:2006ek}, i.e. $\textrm{Re} (S(j)) = S_{\textrm{BH}}$. The second oscillating cosine factor accounts for the sign oscillation of $\Omega_j$ as $j$ varies due to the $(-1)^F$ factor \cite{Agarwal:2020zwm}.

In the above example, the complex conjugate pair of saddle points of the Legendre transformation is to imitate the sign oscillation of $\Omega_j$ of the index by the cosine function. This structure is universal and should be realized in any supersymmetric index on $\mathcal{M}_{D-1} \times S^1$ of the $D$-dimensional unitary QFT, if the index exhibits overall sign oscillation. In particular, one can easily check that the superconformal indices of SCFTs in $D=3$ \cite{Choi:2019zpz,Choi:2019dfu,Bobev:2019zmz,Benini:2019dyp}, $D=4$ \cite{Choi:2018hmj,Benini:2018ywd,Kim:2019yrz,Lezcano:2019pae}, $D=5$ \cite{Choi:2019miv,Crichigno:2020ouj}, and $D=6$ \cite{Choi:2018hmj,Nahmgoong:2019hko} exhibit such complex saddle points of the Legendre transformation at large $N$. (There have been many works on this subject in the recent years. Please refer to the references of the above for further details.)

\subsection{AdS gravity side}
One may want to interpret the above large $N$ complex saddle points and the phase of the entropy in the dual gravity side. Obviously, in the Lorentzian black hole solution, there are no such complex quantities. The Bekenstein-Hawking entropy is just given by the horizon area of the black hole, and the chemical potentials are the electrostatic potentials and the angular velocities at the horizon. So they are all genuinely physical observables. In order to obtain complex quantities, one should Wick rotate the BPS black hole solution, and study the (regularized) Euclidean on-shell action. In \cite{Cabo-Bizet:2018ehj,Cassani:2019mms}, it was shown that the electrically charged rotating BPS black hole solutions in AdS$_{D+1=4,5,6,7}$ become complex as we Wick rotate them to the Euclidean solutions. Furthermore, there are always a complex conjugate pair of Euclidean solutions corresponding to one rotating Lorentzian BPS black hole solution. This conjugate pair of the Euclidean solutions naturally corresponds to the conjugate pair of the Legendre transform saddle points of the CFT index in the large $N$ limit. The extremized chemical potentials, free energy, and the entropy, which are complex, are precisely matched between each solution of the conjugate pair \cite{Cassani:2019mms}.

\section{Complex saddles in 3D-3D correspondence} \label{sec : complex saddles in 3D-3D}
In this section, we consider the universal magnetically charged AdS$_4$ BPS black holes in 4D minimal $\mathcal{N}=2$ gauged supergravity \cite{Romans:1991nq,Caldarelli:1998hg}. Its bosonic action in the Euclidean signature is given in \eqref{SUGRA-action}.
There are two well-known ways of embedding this low energy effective supergravity into a full fledged quantum gravity theory, M-theory. One is using multiple M2-branes probing a cone over Sasaki-Einstein 7-manifold $Y_7$, and the other is using multiple M5-branes wrapped on a compact 3-manifold $M$. The macroscopic constants, $G_4$ and $\theta$, in the 4D supergravity are determined by the microscopic data, $N$ (the number of branes) and geometric quantities of the internal manifold ($Y_7$ or $M$).  The $\theta$-term turns out to vanish, i.e. $\theta=0$, for the 4D supergravity from multiple M2-branes, while it does not vanish for the case of multiple M5-branes. The $\theta$-term has been usually overlooked in the literature, but here we point out that the $\theta$-term naturally appears in the consistent KK truncation of the chiral 7D maximal gauged supergravity \cite{Pernici:1984xx,Gauntlett:2000ng,Donos:2010ax}, which is holographically dual to the 6D (2,0)-theory living on the M5-branes, to the 4D minimal gauged supergravity. Moreover, we demonstrate that the $\theta$-angle is proportional to a topological invariant called the 'Chern-Simons invariant', $cs(M)$, of the internal 3-manifold. 

 The non-trivial $\theta$-term has interesting implications in the large $N$ analysis of the twisted indices of the dual 3D $\mathcal{N}=2$ superconformal field theory. It determines the sign oscillatory behavior of the twisted indices at large $N$. After the Wick rotation, the magnetically charged black holes in AdS$_4$  become  a pair of Euclidean saddle points called Bolt$_{\rm \pm}$ with $p=0$ and $g>1$ \cite{Toldo:2017qsh}.  
  The $\theta$-term contributes as an imaginary part to the Euclidean action  and the on-shell actions for Bolt$_\pm$ are complex and conjugate to each other.   At large $N$, the Bolt$_{\pm}$ solutions give the most dominant contributions to the path integral of the AdS$_4$ quantum gravity with an asymptotic boundary condition associated to the twisted indices. Then, the imaginary part of their on-shell action determine the oscillatory phases factor of the twisted indices at large $N$.  See \eqref{Oscillatory for electrical BHs} for the similar oscillatory phase factor for electrically charged black holes and superconformal indices \cite{Agarwal:2020zwm}. 
  
  The supergravity computation of $\theta$-term also has interesting mathematical implications. Combining  the holographic principle   and 3D-3D relation  for twisted indices, the Euclidean saddles Bolt$_{\rm \pm}$ are mapped to two canonical $SL(N, \mathbb{C})$ flat connections, $\mathcal{A}_N^{\rm geom}$ and  $\mathcal{A}_N^{\overline{\rm geom}}$, on the internal hyperbolic  3-manifold $M$. The on-shell values of  $\theta$-term for Bolt$_{\rm \pm}$ determine the large $N$ behavior of the phase factor of a topological invariant called the adjoint Reidemeister-Ray-Singer torsion associated to the two $SL(N, \mathbb{C})$ flat connections.

\subsection{3D theory $T_N[M]$ from wrapped M5-branes}
Here, we briefly review the 3D  $T_N[M]$ theory geometrically constructed from wrapped $N$ M5-branes on the compact 3-manifold $M$ \cite{Dimofte:2010tz,Dimofte:2011ju,Dimofte:2013iv,Gang:2018wek}.  The theory can be obtained from a twisted compactification of 6D $A_{N-1}$ (2,0)-theory, which is the world-volume theory of $N$ M5-branes:
\begin{align}
\textrm{6D $A_{N-1}$ (2,0)-theory on $\mathbb{R}^{1,2}\times M$} \;\xrightarrow{\qquad \textrm{size}(M)\rightarrow 0  \qquad } \;\textrm{3D $T_{N}[M]$ theory on $\mathbb{R}^{1,2}$}\;.
\end{align}
In the compactification, we perform a partial topological twisting along the compact 3-manifold $M$ to preserve some supersymmetries. In the topological twisting, we use the $SO(3)$ subgroup of 6D $SO(5)$ R-symmetry and the twisting preserves 4 supercharges out of 16. Thus, the resulting  $T_N[M]$ theory has 3D $\mathcal{N}=2$ supersymmetry. Field theoretic construction of the $T_N[M]$ theory is given in \cite{Dimofte:2011ju,Dimofte:2013iv,Gang:2018wek}. From the explict construction,  one can confirm that the $T_N[M]$ only has $U(1)$ R-symmetry and no other flavor symmetry at sufficiently large $N$ for  hyperbolic $M$. The $U(1)$ R-symmetry originates from the $SO(2)$ subgroup of 6D $SO(5)$ R-symmetry and thus the R-charge is integer-quantized:
\begin{align}
R (\mathcal{O} ) \in \mathbb{Z}\;, \; \textrm{for all local operators $\mathcal{O}$ in $T_N[M]$ theory}\;. \label{compact U(1)}
\end{align}
Since there is no  flavor symmetry to mix, the $U(1)$ R-symmetry is identical to the IR superconformal R-symmetry. 

\subsection{3D-3D relations for twisted indices}
One nice feature of the 3D $T_{N}[M]$ theory is that the supersymmetric partition functions of the theory can be written in terms of  invariants of $SL(N,\mathbb{C})$ Chern-Simons theory on $M$. The  simplest 3D-3D relation is the one for twisted indices. The twisted indices of $T_N[M]$ theory on genus $g$ Riemann surface $\Sigma_g$ is defined as
\begin{align}
\begin{split}
&I_{g}[T_N[M]] := \textrm{Tr}_{\mathcal{H} (\Sigma_g)}(-1)^R\;, \quad \textrm{where}
\\
&\mathcal{H} (\Sigma_g)  \;:=\; \textrm{Topologically twisted Hilbert-space  on $\Sigma_g$}\;. \label{Def : Twisted index}
\end{split}
\end{align}
Along $\Sigma_g$, we perform a topological twisting using the compact $U(1)$ R-symmetry. The twisting is  possible only when the following Dirac quantization is satisfied:
\begin{align}
R (\mathcal{O}) \times (g-1)\in \mathbb{Z}\;, \quad \textrm{for all local operators $\mathcal{O}$}\;. 
\end{align}
The condition is always satisfied for $T_N[M]$ theory  thanks to the equation~\eqref{compact U(1)}. The 3D-3D relation for the twisted indices is  \cite{Gang:2018hjd,Gang:2019uay}
\begin{align}
I_{g}[T_N[M]]  = \sum_{\rho \in\chi_{\rm irred}(M;N) } \left(N  \textrm{\bf Tor}_{\rm Adj}\left(\rho_\alpha;SL(N,\mathbb{C})\right)\right)^{g-1}\;. \label{3D-3D for twisted indices}
\end{align}
The above 3D-3D relation holds only for 3-manifold with trivial $H_1(M, \mathbb{Z}_N)$. For  3-manifold with non-trivial $H_1(M, \mathbb{Z}_N)$, the 3D-3D relation should be modified    \cite{Benini:2019dyp,Cho:2020ljj}. But the modification only affects the $\log N$ corrections at large $N$.  
Here $\chi_{\rm irred}(M;N)$ is the set of irreducible $SL(N, \mathbb{C})$ flat connections on $M$
\begin{align}
\chi_{\rm irred}(M;N)   &= \frac{ \{ \rho \in \textrm{Hom}\left(\pi_1 M \rightarrow SL(N, \mathbb{C}) \right) \;:\; \rho \textrm{ is irreducible} \}}{(\rm conjugation)}\;.
\end{align}
The flat connections are classical solutions of $SL(N, \mathbb{C})$ Chern-Simons theory on $M$. 
Conventionally, a  $G$ flat  connection is defined as gauge field configuration $\mathcal{A}$ of  group $G$ with vanishing field strength, i.e. $d \mathcal{A}+ \mathcal{A}\wedge \mathcal{A}=0$. A $G$ flat connection $\mathcal{A}$ induces a homomorphism $\rho : \pi_1 M \rightarrow G$ via a holonomy matrix $\rho(a) := P \exp \left( \oint_a \mathcal{A}\right)$.  The holonomy matrices fully characterize the flat connection and thus we can identify a flat connection $\mathcal{A}$ with its induced homomorphism $\rho$.  Using the  identification, the symbols $\mathcal{A}$ and $\rho$ are interchangeably used throughout this paper. 
A homomorphism $\rho \in \textrm{Hom}(\pi_1 M \rightarrow SL(N, \mathbb{C}))$ is called {\it irreducible} if its stabilizer subgroup $\textrm{\bf Stab}(\rho) \subset SL(N, \mathbb{C})$ defined in the following way is zero-dimensional.
\begin{align}
\textrm{\bf Stab} (\rho) := \{ h \in SL(N, \mathbb{C})\; :\; [h, \rho(a)] =0 \quad \forall a \in \pi_1 M \}\;.
\end{align}
$\textrm{\bf Tor}_{\rm Adj}\left(\rho_\alpha;SL(N, \mathbb{C})\right)$ is a mathematical invariant called the adjoint Reidemeister-Ray-Singer torsion \cite{cheeger1977analytic, reidemeister1935homotopieringe,ray1971r} associated to the flat connection $\rho_\alpha$. The invariant in general representation $R\in \textrm{Hom}[G_\mathbb{C}\rightarrow GL(V_R)]$ can be defined as follows
\begin{align}\label{Def : Torsion}
\textrm{\bf Tor}_{R} [\mathcal{A}_\alpha; G_{\mathbb{C}} ] :=  \frac{[\det' \Delta_1 (R, \mathcal{A}_\alpha)]^{1/2}}{[\det' \Delta_0 (R, \mathcal{A}_\alpha)]^{3/2}}\;.
\end{align}
Here $\Delta_n (R, \mathcal{A}_\alpha)$  is a Laplacian acting on $V_R$-valued $n$-form twisted by a flat connection $\mathcal{A}_\alpha$ and  $\det' $ is the zeta function regularized determinant. In the definition, we need to choose a metric structure on $M$  but the final $\textrm{\bf Tor}_{R}$ turns out to be independent of the choice. 
When $R=({\rm Adj})$ (adjoint representation), the invariant is related to  the perturbative 1-loop  expansion of the complex Chern-Simons theory around the flat connection $\mathcal{A}_\alpha$ in the following way
\begin{align}
\begin{split}
&\int \frac{[d (\delta\mathcal{A})]}{({\rm gauge })} \exp \left(- \frac{1}{2\hbar}  \int_M \textrm{Tr}(\CA d \CA +\frac{2}3 \CA^3)\right)\bigg{|}_{\CA = \CA_\alpha +\delta \CA}
\\
&\xrightarrow{\quad \hbar \rightarrow 0 \quad } \frac{1}{\sqrt{\textrm{\bf Tor}_{\rm Adj}\left(\rho_\alpha;SL(N, \mathbb{C})\right)}} \exp \left(- \frac{1}{2\hbar}  \int_M \textrm{Tr}(\CA_\a d \CA_\a +\frac{2}3 \CA_\a^3) + \CO(\hbar)\right)\;.
\end{split}
\end{align}
As a non-trivial consistency check of the 3D-3D relation, the integrality of the twisted indices in \eqref{3D-3D for twisted indices} were checked for various examples in \cite{Gang:2019uay,Benini:2019dyp,Cho:2020ljj}. 

\subsection{Two dominant $SL(N, \mathbb{C})$ flat connections  at large $N$}
Here, we focus on the case where $M$ is a closed hyperbolic 3-manifold. 
At large $N$, the dominant contributions to the twisted index in \eqref{3D-3D for twisted indices} with $g>1$  come from the two irreducible flat connections, $\rho^{\rm geom}_N$ and $\rho^{\rm \overline{geom}}_N$, related to each other by complex conjugation. The flat connections are defined as follows
\begin{align}
\mathcal{A}^{\rm geom}_N := \tau_N \cdot (\omega + i e)\;, \quad \mathcal{A}^{\overline{\rm geom}}_N := \tau_N \cdot (\omega - i e)\;.
\end{align}
Here $\omega$ and $e$ are spin-connection and dreibein for the unique hyperbolic metric on $M$ normalized as $R_{\mu\nu} = -2 g_{\mu\nu}$. They can be regarded as $so(3)$-valued 1-form on $M$  and their complex combinations $(\omega\pm i e)$ define a complex conjugate pair of $SL(2,\mathbb{C})$ connections.\footnote{More precisely, they are $PSL(2,\mathbb{C})$ flat connections, which can always  be uplifted  to $SL(2,\mathbb{C})$ flat connections.  } The hyperbolicity, $R_{\mu\nu} = -2 g_{\mu\nu}$, implies that the connections are actually flat.  Via the principal embedding  $\tau_N  :SU(2) \rightarrow SU(N)$, the $SL(2,\mathbb{C})$ flat connections are mapped to a complex conjugate pair of $SL(N,\mathbb{C})$ connections,  $\mathcal{A}^{\rm geom}_N $ and $\mathcal{A}^{\overline{\rm geom}}_N$.  At large $N$, the two flat connections give the most dominant contributions to the  twisted indices in \eqref{3D-3D for twisted indices} with $g>1$, i.e.
\begin{align}
\begin{split}
I_{g>1}[T_N[M]]   \xrightarrow{\quad N \rightarrow \infty \quad } & \left( N  \textrm{\bf Tor}_{\rm Adj}(\rho^{\rm geom}_N ;SL(N,\mathbb{C}))\right)^{g-1} +(c.c)
\\
& + (\textrm{exponentially smaller terms})\;. \label{large N limit: two saddles}
\end{split}
\end{align}
We now study the large $N$ limit of the adjoint torsion $ \textrm{\bf Tor}_{\rm Adj}(\rho^{\rm geom}_N ;SL(N,\mathbb{C}))$. First, from the following branching rule
\begin{align}
(\textrm{Adjoint representation of }SU(N)) = \left( \bigoplus_{n=1}^{N-1} \tau_{2n+1} \textrm{ of }SU(2)\right)\;,
\end{align}
we have
\begin{align}
\log \textbf{Tor}_{\rm Adj} \left(\rho^{\rm geom}_N;SL(N, \mathbb{C})\right) = \sum_{n=1}^{N-1} \log \textbf{Tor}_{\tau_{2n+1}} \left(\rho^{\rm geom}_{N=2};SL(2, \mathbb{C})\right) \;. \label{branching rule 2}
\end{align}
In the above,  the $SU(2)$ is embedded into $SU(N)$ via the principal embedding $\tau_N$ and $\tau_{2n+1}$ is the $(2n+1)$-dimensional irreducible representation of $SU(2)$.  $\textrm{\bf Tor}_{\tau}\left(\rho;SL(2, \mathbb{C})\right)$ is the  Reidemeister-Ray-Singer torsion associated to the  $SL(2,\mathbb{C})$ flat connection $\rho$ in the representation $\tau$. Using Selberg's trace formula, it is mathematically proven that
\begin{align}
\begin{split}
&\textbf{Theorem \cite{park2017reidemeister}: }
\\
&\textrm{Re} \left( \log \textbf{Tor}_{\rm \tau_{2n+1} }(\rho_{N=2}^{\rm geom}, SL(2, \mathbb{C})) \right) = \frac{1}{\pi} \textrm{vol}(M) \left(n^2+n+\frac{1}6\right) + \textrm{Re}\sum_{\gamma}\sum_{k=n+1}^{\infty} \log (1-q_\gamma^k)\;. \label{Selberg-formula}
\end{split}
\end{align}
Here ${\rm vol}(M)$ is the hyperbolic volume, volume measured using the unique hyperbolic metric normalized as $R_{\mu\nu } = -2 g_{\mu\nu}$.  In the last term,  the summation is over geodesics $\gamma$ in $M$ and $q_\gamma := e^{-\ell_{\mathbb{C}}(\gamma)}$ where $\ell_{\mathbb{C}}(\gamma)$ is the complex length of $\gamma$. The complex length is defined as
\begin{align}
\textrm{Tr}(\rho^{\rm geom}_{N=2} (\gamma)) = 2 \cosh \left( \frac{\ell_{\mathbb{C}}(\gamma)}2 \right)\;, \quad \textrm{Re} (\ell_{\mathbb{C}})>0\;.
\end{align}
Combining \eqref{branching rule 2} and \eqref{Selberg-formula}, we have the following large $N$ behavior of the adjoint torsion \cite{Gang:2019uay}. 
\begin{align}
\begin{split}
&\textrm{\bf Tor}_{\rm Adj}(\rho^{\rm geom}_N ; SL(N, \mathbb{C}) ) 
\\
&= e^{i \varphi(N, M)}  \exp \left( \frac{\textrm{vol}(M)}{6\pi } (2N^3-N-1) +  \textrm{Re} \sum_{[\gamma]} \log \textrm{P.E.}\left[\frac{q_\gamma^{N+1}-q_\gamma^2}{(1-q_\gamma)^2}\right] \right)\;. \label{large N of torsion}
\end{split}
\end{align}
 Here ${\rm P.E.}$ denotes the plethystic exponential,
 \begin{align}
 {\rm P.E.}[f(q)] := \exp \left(\sum_{n=1}^\infty \frac{1}{n} f(q^n)\right)\;.
 \end{align}
 Then, using \eqref{large N limit: two saddles}, we finally have following large $N$ behavior 
\begin{equation} 
\begin{aligned}
I_g [T_N[M]] &\sim \exp \left( (g-1)\frac{\textrm{vol}(M)}{6\pi}(2 N^3-N) + i(g-1) \varphi (N, M)+ o(N) \right) + (c.c.) \\
& = \exp \left( (g-1)\frac{\textrm{vol}(M)}{6\pi} (2N^3-N) + o(N) \right)\times \big{(} 2 \cos \left( (g-1)\varphi(N, M) \right) \big{)}\ .
\end{aligned}
\end{equation}
The exponential growth of $I_{g}[T_N[M]]$ at large  $N$  nicely matches with the Bekenstein-Hawking entropy of the corresponding magnetically charged black hole in the gravity dual including subleading corrections \cite{Gang:2018hjd,Gang:2019uay,Benini:2019dyp,Bobev:2020egg,Bobev:2020zov}.  

In this paper,  we focus on  the oscillatory factor, $2 \cos \left( (g-1)\varphi(N, M) \right)$, of the large $N$ twisted indices.  As the main result, from holographic dual computation in the next section,  we propose that
\begin{align}
\varphi (N, M) \xrightarrow{ \quad N \rightarrow \infty \quad } \left( \frac{cs (M)}{3\pi } N^3 + \CO (N)\right) \quad (\textrm{mod }2\pi)\;. \label{main proposal}
\end{align} 
Here, $cs(M)$ is the so-called Chern-Simons invariant of the hyperbolic 3-manifold $M$, which  is defined as follows
\begin{align}
\frac{i}2 \int_M  \textrm{Tr} \left( \mathcal{A}^{\rm geom}_{N=2} \wedge d \mathcal{A}^{\rm geom}_{N=2}   + \frac{2}3 \mathcal{A}^{\rm geom}_{N=2} \wedge \mathcal{A}^{\rm geom}_{N=2} \wedge \mathcal{A}^{\rm geom}_{N=2}\right) = \textrm{vol}(M) + i cs(M)\;.
\end{align}
The invariant $cs(M)$ is defined only modulo $\pi^2$.   The above holographic prediction combined with the 3D-3D relation \eqref{3D-3D for twisted indices} implies the following non-trivial mathematical conjecture
\begin{align}
\begin{split}
&\textbf{Conjecture : }
\\
&\textrm{Im} \left( \log \textbf{Tor}_{\rm \tau_{2n+1} }(\rho_{N=2}^{\rm geom}, SL(2, \mathbb{C})) \right)  \xrightarrow{\quad n\rightarrow \infty \quad  } \frac{1}{\pi} cs(M) (n^2+ n ) + \CO(n^0)  \quad  \left( \textrm{mod}  \;2\pi \right)\;. \label{conjecture}
\end{split}
\end{align}
Combined with the formula in \eqref{branching rule 2}, the conjecture implies the proposal in \eqref{main proposal}. We will confirm numerically the conjecture with an explicit example in Appendix \ref{App : numerical check}.

\subsection{Supergravity analysis}
The holographic dual of $T_N[M]$ for a closed hyperbolic 3-manifold $M$ is proposed in \cite{Gauntlett:2000ng}. At low energy, the gravity theory is described by the 4D minimal $\mathcal{N}=2$ gauged supergravity whose Euclidean action is given in \eqref{SUGRA-action} with 
%
\begin{align}
L^2/G_4  = \frac{2(N^3+\CO(N)) \textrm{vol}(M)}{3\pi^2}\;.
\end{align}
The 4D supergravity action can be obtained from a consistent KK truncation of 7D maximal gauged supergravity \cite{Pernici:1984xx} on the 3-manifold $M$. 
In the consistently truncated 4D supergravity, there is a  following topological term, which has been overlooked in the literature 
\begin{align}
S_{\theta\textrm{-term}} = - i  \frac{N^3 + \CO(N)}{6 }\times  \frac{cs(M)}{2\pi^3 L^2} \times \int F\wedge F\;, \;\; \; \textrm{i.e.}  \;\; \theta = -  \frac{N^3+ \mathcal{O}(N)}{3} \frac {cs(M)}{\pi}\ . \label{theta-term}
\end{align}
Note that $cs(M)$ is only defined  modulo $\pi^2$ and $ \int F\wedge F \in 4\pi^2 L^2 \mathbb{Z}$.\footnote{The bulk gauge field $A$ is normalized as $A= i L A_{\rm CFT}$, where $A_{\rm CFT}$ is the background gauge field for the compact $U(1)$ R-symmetry in the conformal field theory at the boundary. In  our convention,  the boundary $A_{\rm CFT}$ is purely imaginary.  The R-symmetry is normalized so that $R(Q)  = \pm 1$ for  supercharges $Q$.  } For the $\exp \left(- S_{\theta \textrm{-term}}\right)$ to be well-defined, we require that
\begin{align}
\frac{N^3 + \CO(N)}{6 } \in \mathbb{Z}\;.
\end{align}
This gives a non-trivial consistency condition for the $\CO(N^1)$ correction in $\theta$. One natural guess is  
\begin{align}
\theta = - \frac{N^3-N}{3} \frac {cs(M)}{\pi} \;.
\end{align}
We are going to evaluate the topological $\theta$-term on the $\textrm{Bolt}_\pm$ solutions with $g>1$ and $p=0$. They are Euclidean supergravity saddles obtained after Wick rotating the magnetically charged AdS$_4$ black holes \cite{Toldo:2017qsh}.  
 Topologically, the Euclidean solutions satisfy 
\begin{align}
\begin{split}
&\mathcal{M}^{\textrm{Bolt}_\pm}_4  \sim   \Sigma_g \times D_2\;,
\\
&F^{\textrm{Bolt}_\pm} \sim  \frac{L}2\left(\textrm{vol} (\Sigma_g) \mp  \textrm{vol}(D_2)\right)\;. \label{global structure of bolt}
\end{split}
\end{align}
Refer to Appendix \ref{Appendix : Bolt} for details. 
Here, the volume form are normalized as
\begin{align}
\begin{split}
&\int_{\Sigma_g} \textrm{vol}(\Sigma_g) = 4\pi  (1-g)\;,
\\
&\int_{S^2 = D_2 \bigcup D_2 } \textrm{vol}(D_2) = 2 \times \int_{D_2  } \textrm{vol}(D_2)  = 4\pi  .  \;
\end{split}
\end{align}
At the boundary $\partial (\mathcal{M}_4^{\textrm{Bolt}_\pm }) = \Sigma_g\times S^1$, the gauge field configuration is given by
\begin{align}
F^{\textrm{Bolt}_\pm}_{\rm CFT}=\frac{1}{2i} \textrm{vol}(\Sigma_g) \mp  \frac{1}{2i} \textrm{vol}(D_2)\; . \nonumber
\end{align}
The first term is compatible with the fact that there is topological twisting on $\Sigma_g$ and the second term is compatible with the factor $(-1)^R$ in the definition of the twisted index \eqref{Def : Twisted index} since
\begin{align}
\quad \exp \left( \oint_{S^1} A_{\rm CFT}^{\textrm{Bolt}_\pm }\right)= \exp \left( \int_{D_2} F_{\rm CFT}^{\textrm{Bolt}_\pm }\right) =\exp (\pm  i \pi ) = -1\;.
\end{align}
 In the Euclidean path integral of the twisted index, we impose anti-periodic boundary condition for fermionic fields, which is also compatible with the shrinking of the $S^1$-cycle in the bulk. 

Using \eqref{global structure of bolt}, we finally obtain
\begin{align}
\begin{split}
S_{\theta\textrm{-term}} [\textrm{Bolt}_\pm ] &= - i  \frac{N^3+\mathcal{O}(N)}{6}\times  \frac{cs(M)}{2\pi^3 L^2} \times 2 \times \left( \int_{\Sigma_g} \frac{L}2 \textrm{vol}(\Sigma_g) \right)\times \left(\mp   \frac{L}2  \int_{D_2} \textrm{vol}(D_2) \right) 
\\
& =  \pm  i    \frac{N^3+\CO(N)}{3\pi} (1-g) \times  cs(M)\;.
\end{split}
\end{align}
The above computation then holographically implies the proposal in \eqref{main proposal}.

\acknowledgments

We thank Nikolay Bobev, Anthony M. Charles, Kiril Hristov, Jinsung Park, Valentin Reys  and especially Seok Kim for insightful comments and discussions. SC thanks APCTP for kind hospitality during his visiting, where part of this work was done. SC is partially supported by the National Research Foundation of Korea (NRF) Grant 2018R1A2B6004914 and by NRF-2017-Global Ph.D. Fellowship Program.
The research of DG and NK is supported by the NRF Grant 2019R1A2C2004880. DG also acknowledges support by the appointment to the JRG program at the APCTP through the Science and Technology Promotion Fund and Lottery Fund of the Korean Government, as well as support by the Korean Local Governments, Gyeongsangbuk-do Province, and Pohang City.

\appendix
\section{Global structure of ${\rm Bolt}_\pm$ solutions} \label{Appendix : Bolt}
For the computation of the topological  $\theta$-term in \eqref{theta-term}, we consider a non-supersymmetric  deformation of the ${\rm Bolt}_\pm$ solutions with $p=0$, where the topological structure becomes more manifest.
The deformed Euclidean background for $g>1$ is given as follows \cite{Toldo:2017qsh,Bobev:2020zov}
\begin{align}
\begin{split}
&ds^2/L^2 = \lambda (\rho) \left(d \tau -2s \cosh \theta d\phi \right)^2 + \frac{d\rho^2}{\lambda(\rho)} + (\rho^2-s^2) ds^2 (\Sigma_g)\;,
\\
&A_\tau = \frac{L}{2s}\frac{P(\rho^2+s^2)-2s Q\rho}{\rho^2-s^2}\;, \; A_\phi =-  L\frac{P(\rho^2+s^2)-2s Q\rho}{\rho^2 - s^2} \cosh \theta \;.
\end{split}
\end{align}
Here we define
\begin{align}
 \lambda(\rho) := \frac{(\rho^2-s^2)^2+(-1-4s^2)(\rho^2+s^2)-2M\rho +P^2-Q^2}{\rho^2 -s^2}\;.
\end{align}
The supersymmetric  ${\rm Bolt}_\pm$ solutions with $p=0$ correspond to 
\begin{align}
\begin{split}
&P= -2s^2 - \frac{1}2\;, \quad M=2sQ_{\pm}\;, \quad Q= Q_{\pm}\;, 
\\
&\textrm{with }Q_{\pm} = \mp \frac{16s^2 \sqrt{(16s^2)^2+128 s^2}}{128 s^2}\;.
\end{split}
\end{align}
We consider the deformed geometry with $s= \epsilon,  M=0, P=-\frac{1}2 $ with small $\epsilon>0$. When $\epsilon\rightarrow  0$ and $Q\rightarrow \mp 0$, the deformed geometries approach to the ${\rm Bolt}_{\pm}$  geometries. In the limit, the deformed geometry  simplified as 
\begin{align}
\lambda (\rho) = \rho^2 -1 +\frac{\frac{1}4 -Q^2}{\rho^2} + o(\epsilon^2)\;, \quad F_{\rho \tau}  = \frac{LQ}{\rho^2} +o (\epsilon)\;, \quad F_{\Sigma_g} =\frac{L}2 \textrm{vol}(\Sigma_g) +o (\epsilon)\;.
\end{align}
Note that $\rho \geq \rho_0 := \frac{\sqrt{1+2 |Q|}}{\sqrt{2}}$. To avoid conical singularity at $\rho=\rho_0$, $\tau$ should be periodic with the following periodicity
\begin{align}
\Delta \tau = \frac{4\pi}{|\lambda'(\rho_0)|} =  \frac{\pi  \sqrt{|Q|+\frac{1}{2}}}{|Q|}\;.
\end{align}
So the topology of the deformed geometry is $\mathcal{M}^{{\rm Bolt}_\pm }_4\sim D_2\times \Sigma_g$, where $\rho$ parametrizes the radial direction of $D_2$. Along the $\Sigma_g$, magnetic flux $\textrm{vol}(\Sigma_g)$ is turned on. To measure the flux along $D_2$, we compute
\begin{align}
\int_{D_2} F = \left( \int_{\rho_0}^\infty \frac{LQ}{\rho^2}  \right) \times \Delta \tau = \frac{L Q}{\sqrt{|Q|+\frac{1}2}} \times      \frac{\pi  \sqrt{|Q|+\frac{1}{2}}}{|Q|} = \mp  \pi L \;. 
\end{align}
Due to the topological property of the integral, the computation is exact in $\epsilon$. Thus, it implies that $F_{D_2} \sim  \mp \frac{L}2  \textrm{vol}(D_2)$. 
 
\section{Numerical verification of the conjecture \eqref{conjecture} with $M=(S^3\backslash \mathbf{4}_1)_{P/Q}$} \label{App : numerical check}
Let $(S^3\backslash \mathbf{4}_1)_{P/Q}$ be a 3-manifold obtained by a Dehn surgery along the figure-eight knot (denoted by $\mathbf{4}_1$) with a slope $P/Q \in \mathbb{Q}$. The manifold is hyperbolic except $P/Q \in \{0, \pm 1, \pm 2, \pm 3, \pm 4 \}$.
Fundamental group of the manifold is given as follows
\begin{align}
\pi_1 M = \langle a, b, \mathbf{m}, \mathbf{l} \;:\;  ab^{-1}a^{-1} b a = b a b^{-1} a^{-1}b, \; \mathbf{m}=a,\; \mathbf{l} = ab^{-1} a b a^{-2} b a b^{-1}a^{-1},\; \mathbf{m}^P \mathbf{l}^Q=1 \rangle \;. \nonumber
\end{align} 
For a given irreducible $SL(2,\mathbb{C})$ flat connection $\rho \in \textrm{Hom}(\pi_1 M \rightarrow SL(2,\mathbb{C}) )$, its Ray-Singer-Reidemeister torsion $\textbf{Tor}_{\tau_{2n+1}}\left(\rho, SL(2,\mathbb{C})\right)$ can be computed as follows \cite{hiroshi2017twisted}
\begin{align}
\begin{split}
&\textbf{Tor}_{\tau_{2n+1}}\left(\rho, SL(2,\mathbb{C})\right)
\\
&=\left( \frac{P(\ell - \frac{1}{\ell}) m^4 }{(m^4 - 1)(4-2 m^2 +4 m^4)} + Q \right)  \frac{ \lim_{t\rightarrow 1} (t-1)^{-1}\Delta (t;\tau_{2n+1},\rho) }{\prod_{a=1}^n (1-m^{2aR}\ell^{2aS} )(1-m^{-2aR}\ell^{-2aS} )} \;,
\\
&\textrm{with } \Delta (t;\tau_{2n+1},\rho) := \frac{\det(I_{2n+1} -t^{-1} A_n B_n^{-1} A_n^{-1} +A_n B_n^{-1}A_n^{-1} B_n - t B_n +B_n A_n B_n^{-1}A_n^{-1})}{\det (tI_{2n+1}-B_n)}\;.
\end{split}
\end{align}
Here $A_n:= \tau_{2n+1}(\rho(a)), B_n:=\tau_{2n+1}(\rho(b))$ and $I_{2n+1}$ is the identity matrix of $(2n+1)$. $m$ and $\ell$  are defined in the following relation
\begin{align}
\bigg{(} \rho(\mathbf{m}) ,  \rho(\mathbf{l})  \bigg{)}\sim_{\rm conj}  \left(  \begin{pmatrix}
m & * \\
0 & m^{-1} 
\end{pmatrix}, \begin{pmatrix}
\ell & * \\
0 & \ell^{-1} 
\end{pmatrix}  \right) \;.
\end{align}
Here, two integers $R$ and $S$ are chosen such that $PS-QR=1$. 
\paragraph{$M=(S^3\backslash \mathbf{4}_1)_{P/Q=5}$ case } The holonomy matrices for the flat connection $\rho^{\rm geom}_{N=2}$ are \footnote{There are only 4 irreducible $SL(2,\mathbb{C})$ flat connections on the 3-manifold. Two of them are real, i.e. trace of all holonomy matrices are real, and the other twos correspond to  $\mathcal{A}^{\rm geom}_{N=2}$ and $\mathcal{A}^{\overline{\rm geom}}_{N=2}$. The notion of being $\mathcal{A}^{\rm geom}_{N=2}$ or $\mathcal{A}^{\overline{\rm geom}}_{N=2}$ depends on the orientation choice of $M$ and we in particular choose one choice.  }
\begin{align}
\begin{split}
&\rho^{\rm geom}_{N=2}(a) = \left(
\begin{array}{cc}
0.58480+0.37948 i & 0 \\
-1 & 1.2033-0.7808 i \\
\end{array}
\right)\;,
\\
&\rho^{\rm geom}_{N=2}(b) = \left(
\begin{array}{cc}
0.58480+0.37948 i & -0.99245+0.51312 i \\
0 & 1.2033-0.7808 i \\
\end{array}
\right)\;.
\end{split}
\end{align}
Then, its Ray-Singer-Reidemeister  torsions are
\begin{align}
\begin{split}
&\{ \log \textbf{Tor}_{\tau_{2n+1}} [\rho^{\rm geom}_{N=2}] \}_{n=1}^{15}  \quad  (\textrm{mod } 2\pi i  )
\\
&= 
\{1.66019-2.42623 i,\;2.47618+1.40108 i,\;4.36609-2.48725 i,\;6.81514+1.25657 i,
\\
&\qquad 9.90447-0.16483 i,\; 13.75025-0.61606 i,\; 18.0928-0.1422 i,\; 23.0814+1.3424 i\}
\\
&\qquad  28.7128-2.5184 i,\; 34.9561+0.8746 i,\; 41.8289-1.0394 i,\; 49.3278-1.9892 i,
\\
&\qquad  57.4486-1.9708 i,\; 66.1951-0.9834 i,\;75.5668+0.9714 i\}\;.
\end{split}
\end{align}
The above numerical values are compatible with the theorem in \eqref{Selberg-formula} and the conjecture in  \eqref{conjecture} since the following series well converges:
\begin{align}
\begin{split}
&\big{\{} \log \textbf{Tor}_{\tau_{2n+1}} [\rho^{\rm geom}_{N=2}] - \frac{\textrm{vol}(M)}{\pi} \left(n^2+n+\frac{1}6\right) -    i \frac{cs(M)}{\pi} \left(n^2+n\right) \big{\}}_{n=1}^{15}  \quad  (\textrm{mod } 2\pi i  )
\\
&=  \{0.983369 +2.88886 i,\; 0.549836 +4.77999 i,\; 0.56547 +4.27058 i,\; 0.515488 +4.14204 i,
\\
&\qquad 0.481026 +4.16337 i, \; 0.578252 +4.18678 i,\; 0.547515 +4.16725 i , \; 0.538042+4.19027 i \}
\\
&\qquad 0.546595 +4.18297 i,\; 0.542266+4.17826 i,\; 0.542725 +4.18166 i,\; 0.544512 +4.18118 i,
\\
&\qquad 0.543498 +4.18071 i,\; 0.543412 +4.18129 i,\; 0.543708 +4.18107 i\} \;.
\end{split}
\end{align}
Here we used the fact that $\textrm{vol}(M) = 0.98137$ and $cs(M) = 1.52067$ \cite{SnapPy}. 

In the same way, we also  numerically confirmed the conjecture for other examples of hyperbolic $M=(S^3\backslash \mathbf{4}_1)_{P/Q}$.
\bibliographystyle{ytphys}
\bibliography{ref}

\end{document}